\documentclass[english,reqno12pt]{iopart}
\usepackage[T1]{fontenc}
\usepackage[letterpaper]{geometry}
\usepackage{amsthm}
\usepackage{amstext}
\usepackage{amssymb}
\usepackage{esint}
\usepackage{textcomp}
\usepackage[dvips]{graphicx} 
\usepackage{babel}

\providecommand{\tabularnewline}{\\}
\providecommand{\be}{\begin{equation}}
\providecommand{\ee}{\end{equation}}
\providecommand{\bea}{\begin{eqnarray}}
\providecommand{\eea}{\end{eqnarray}}
\providecommand{\beas}{\begin{eqnarray*}}
\providecommand{\eeas}{\end{eqnarray*}}
\makeatletter
\usepackage{bbm}
\makeatother
\usepackage{babel}
\begin{document}
\title{Adversarial Satisfiability Problem}

\author {Michele Castellana$^{1,2,3}$, Lenka Zdeborov\'a$^{3,4}$}
\address{$^ 1$Dipartimento di Fisica, Universit\`a di Roma `La Sapienza' , 00185 Rome, Italy} 
\address{$^ 2$LPTMS, CNRS and Universit\'e Paris-Sud, UMR8626, B\^{a}t. 100, 91405 Orsay, France}
\address{$^ 3$Theoretical Division and Center for Nonlinear Studies, Los Alamos National Laboratory, NM 87545 USA}
\address{$^ 4$Institue de Physique Th\'eorique, IPhT, CEA Saclay, and URA 2306, CNRS, 91191 Gif-sur-Yvette cedex, France}

\ead{michele.castellana@lptms.u-psud.fr, lenka.zdeborova@cea.fr}

\begin{abstract}
We study the adversarial satisfiability problem, where the adversary can choose whether variables are negated in clauses or not in order to make the resulting formula unsatisfiable. This is one case of a general class of adversarial optimization problems that often arise in practice and are algorithmically much harder than the standard optimization problems. We use the cavity method to compute large deviations of the entropy in the random satisfiability problem with respect to the negation-configurations. We conclude that in the thermodynamic limit the best strategy the adversary can adopt is extremely close to simply balancing the number of times every variable is and is not negated. We also conduct a numerical study of the problem, and find that there are very strong pre-asymptotic effects that are due to the fact that for small sizes exponential and factorial growth is hardly distinguishable.      
\end{abstract}


\date{\today}

\maketitle

\section{Introduction}\label{sec_intro}

The following setting often arises in practical optimization problems. Consider two players, each of them has a given set of moves (configurations) and a cost function depending on the moves of both the players. The first player is trying to optimize a certain cost function over his set of moves (configurations), and the interest of the second player is to make this optimum the worse possible. Consider that in the game at first the second player (adversary) chooses his moves (a configuration), and then the first player chooses his moves. What is the best strategy (algorithm) for the adversary in the case his set of moves is too large to be able to evaluate all the possibilities? A specific example of this {\it adversarial optimization} setting could be a police department trying to set border controls in such a way that the amount of goods smugglers can transfer is the smallest possible \cite{McMastersMustin70,Wood93}. 

Let us call the set of moves of the adversary $\vec u$, and the set of moves of the first player $\vec v$, the cost function being $f(\vec u, \vec v)$. The goal of the adversary is to find $u_m$ that maximizes $\min_{\vec v} f(\vec u, \vec v)$. Since in common situations both $\vec u$ and $\vec v$ have exponentially many components in the size of the system the adversarial optimization is much harder than usual (one-player) optimization, because even evaluating the $\min_{\vec v} f(\vec u, \vec v)$ for a given $\vec u$ is typically an NP-hard problem. In the theory of algorithmic complexity the so-called NP problems are such for which it is easy (polynomial) to evaluate if a proposed solution is indeed a solution. In other words verifying a solution to an NP problem is a polynomial problem. Verifying a solution for a general adversarial optimization problem (as described above) is itself an NP problem. 
Without doubts theoretical understanding of hard optimization problems is crucial for many areas of science, and the same holds for the adversarial optimization. 

In optimization, the most famous benchmark of a hard problem is the $K$-satisfiability ($K$-SAT) of Boolean formulas. Call a {\it clause} a logical disjunction (operation ``or'') of $K$ variables or their negations. Given a set of $N$ Boolean variables $x_i$, and a set of $M$ clauses, the satisfiability problem consists in deciding whether all clauses can be simultaneously satisfied. The $K$-SAT problem was the first problem shown to be NP-complete, that is as hard as any other NP problem \cite{Cook71}. It has a large number of applications in automated verification and design. Random $K$-SAT problem, where variables in clauses are chosen randomly and negated with probability $1/2$, provides easy to generate hard formulas \cite{MitchellSelman92}.

Random $K$-SAT thus became a common playground for new algorithms, and theoretical ideas for understanding the origin of algorithmic hardness. Statistical physics approach related to physics of diluted spin glasses contributed tremendously to the understanding of the properties of random $K$-SAT formulas, see e.g. \cite{MezardParisi02,KrzakalaMontanari06}. Following this path in this paper we introduce a random adversarial satisfiability problem and
develop a statistical mechanics framework to understand its properties. This framework can be readily applied to other random adversarial optimization problems. We study the large deviation functions for the original optimization problem with respect to the moves of the adversary. The main ideas of our approach to the study of large deviations come from studies of spin glasses and the cavity method \cite{Rivoire05,ParisiRizzo08,ParisiRizzo09,ParisiRizzo10}.  Our approach is also closely linked to the well established fact that the replicated free energy in Parisi's replica symmetry breaking (RSB) \cite{Parisi80,DotsenkoFranz94} can be interpreted as Legendre transformation of the large deviation function. We will, however, derive our method independently of these notions, only using the factor graph representation of the problem and belief propagation (BP) algorithm. 

One natural setting for random adversarial satisfiability problem is to introduce the negation-variables $J_{ia}$, where $J_{ia}=1$ if variable $i$ is negated in clause $a$, and $J_{ia}=0$ if not. The set of moves for the adversary are then all the possible configurations of negations $\{J_{ia}\}$, the moves of the second player are all possible configurations of the variables $\{x_i\}$. The graph of interactions is chosen at random as before. The goal of the adversary is to set the negations in such a way that the resulting formula is the most frustrated possible. In particular we will be interested in the question: can the adversary  make the formula unsatisfiable or not? We will call this problem random AdSAT. 

An independent interest in the random AdSAT comes from the study of random quantum satisfiability problem \cite{LaumannMoessner09,BravyiMoore09,LaumannLauchli10}. It was shown that if the adversary can make the formula unsatisfiable then also the quantum SAT is unsatisfiable. A natural question is then whether the quantum SAT is yet much more restrictive than the AdSAT or not? 

Note also that the quantified satisfiability (QSAT) problem is another SAT-based problem that naturally belongs to the general adversarial optimization setting. In QSAT problem we introduce two types of variables the existential variables $x_i$ and the universal variables $y_i$ the QSAT then consists in deciding whether $\forall \vec y \, \,  \exists \vec x \, \,  G(\vec x,\vec y)={\rm TRUE}$ 
%(\textbf{What does this formula mean??}) Is it better with the ={\rm TRUE} ?
, where $G(\vec x,\vec y)$ is a satisfiability formula (with negation-variables fixed). 
Random ensembles of QSAT were introduced and studied \cite{ChenInterian05}. QSAT is arguably more important for industrial applications than the AdSAT that we study here. We chose the AdSAT defined as above because it is slightly simpler to treat and also provides information relevant to the original random SAT problem. We plan to apply our approach to the random QSAT problem in near future. Our study is also related to the work on optimization under uncertainty \cite{AltareliBraunstein10}.  

The present article is structured  as follows: In Section \ref{sec_dev} we set the problem of adversarial satisfiability, and describe our approach in generic statistical physics terms. In Section \ref{sec_BP} we remind the standard belief and survey propagation (SP) equations for random satisfiability problem. In Section \ref{sec_deriv} we derive equations for calculation of the large deviations with respect to the negation-configurations. In Section \ref{sec_res} we first present and discuss the cavity result for random regular adversarial satisfiability, then we do the same for the canonical (Poissonian) random adversarial satisfiability. In Section \ref{sec_num} we compare our theoretical result to numerical simulation that include exhaustive search of all the solutions. Finally in Section \ref{sec_concl} we conclude and discuss perspectives of this work. 

\section{AdSAT as a large deviation calculation}\label{sec_dev}

The random $K$-SAT problem is defined as follows, consider $N$ Boolean variables $\left\{x_i\right\}_{i=1,\cdots,N}$, $x_i=\{0,1\}$, and $M=\alpha N$ clauses $\psi_a$. Each clause depends on $K$ random variables from the $N$ available ones. If a variable $i$ belongs to clause $\psi_a$, then we set $J_{ai}=1$ if the variable is negated, and $J_{ia}=0$ if it is not. $K$-SAT problem can be represented via a so-called factor graph, a bipartite graph between variables (variables nodes) and clauses (function nodes). With edges between variables $i$ that belong to clause $a$. The negation-variables can be seen as attributes of the edges. The random $K$-SAT instance corresponds to the case where  the $J_{ia}$s are drawn uniformly at random. Probably, the most well known property of the random $K$-SAT is the existence of a phase transition at a value $\alpha_c$ such that if $\alpha<\alpha_c$ then with high probability (probability going to one as $N\to \infty$) there exists a configuration $\{x_i\}$ that satisfies all the clauses, and for $\alpha>\alpha_c$ no such configuration exists with high probability. We  define $\partial_i$ as the ensemble of function nodes connected to the variable node $i$,  $\partial_a$ as the ensemble of the variable nodes connected to the function node $\psi_a$.

The adversarial satisfiability problem (AdSAT) is defined by drawing a random $K$-SAT instance as before without deciding the negation-variables $\{ J_{ia} \} \equiv \mathcal{J}$.
A solution to the AdSAT problem is given by a set $\mathcal{J}$ such that the resulting instance is unsatisfiable. Just as in random $K$-SAT there is a threshold $\alpha_a$ in the random AdSAT such that for $\alpha<\alpha_a$ no solution to the AdSAT formula exist with high probability. And for $\alpha>\alpha_a$ a solution exists with high probability. We observe $\alpha_a\le \alpha_c$ since above $\alpha_c$ a random configuration of negations makes the formula unsatisfiable, recall $\alpha_c(K=3)=4.2667$ \cite{MertensMezard06}. Also $\alpha_a\ge \alpha_p$, where $\alpha_p=1/K$ is the percolation threshold below which the graph is basically a collection of small trees and few single loop components, which are both satisfiable for any configuration of negations. One of the goals of the present paper is to estimate the value of the AdSAT threshold $\alpha_a$.

In random $K$-SAT the satisfiability threshold can be found by counting the number of configurations that have a certain energy $E(\{x_i\})$ (i.e. number of unsatisfied clauses). To compute the entropy one introduces a Legendre parameter $\beta$ and computes the free energy $f$ defined as 
\be
          e^{-\beta N f(\beta)} = \sum_{\{x_i\}} e^{-\beta E(\{x_i\})}  = e^{N[s(e)-\beta e]}\, , \quad \quad   \frac{\partial s(e)}{\partial e} = \beta \, ,
\ee
where the number of configurations having energy $E$ is $e^{S(E)}$. If $E=0$ belongs to the support of the function $S(E)$ then the problem is in the satisfiable phase, if not than the problem is in the unsatisfiable phase. In the satisfiable phase call $s=S(0)/N$ the entropy of satisfying configurations. The cavity method and the replica symmetry breaking serve to compute $f(\beta)$ in the thermodynamic limit $N\to \infty$ \cite{MezardParisi01,MezardParisi02}. There are two crucial properties that make this calculation possible: First, the energy can be written as a sum of local terms
\be
   E(\{x_i\}) = \sum_a  \prod_{i\in \partial a} \delta_{x_i,J_{ia}}
\ee
Second, the underlying factor graph is locally tree-like. These computations moreover provide much more information about the problem than the value of the satisfiability threshold. 

In the study of random AdSAT we will proceed analogously. We consider the number of configurations of the negations that yield a given value of the entropy of solutions $s$ 
\be
         s(\mathcal{J}) = \frac{1}{N} \log{\left[ \sum_{\{x_i\}} \prod_{a=1}^M \left(1- \prod_{i\in \partial a} \delta_{x_i,J_{ia}}\right)  \right]}\, .
\ee
We define a large deviation function ${\cal L}(s)$ as the logarithm of this number divided by the size of the system $N$. Again to compute ${\cal L}(s)$ it is advantageous to introduce its Legendre transform 
\be
   Z(x)=  e^{N\Phi(x)} = \sum_{\mathcal{J}} e^{x N s(\mathcal{J})} = e^{N[{\cal L}(s) + xs]}\, , \quad \quad  \frac{\partial {\cal L}(s)}{\partial s} = -x \, . \label{LL}
\ee
We stress here that ${\cal L}(s)$ is the large deviation function with respect to the negation-configurations, it it hence defined for a given geometry of the satisfiability formula. In what follows we assume, as is usual, that ${\cal L}(s)$ is self-averaging with respect to the formula geometry, i.e. ${\cal L}(s)$ is almost surely the same function for two randomly chosen formulas. This assumption is a generalization of the self-averaging property of the free energy in the canonical $K$-SAT problem. Note also that in writing this expression we implicitly assume that the number of negation-configurations that give a certain entropy is exponential in $N$. If it is smaller that exponential in $N$ computation of $\Phi(x)$ will lead to ${\cal L}(s)=-\infty$. We will come back to this point in Sec.~\ref{sec_num}. 

Remark two special cases: For $x=0$ the partition function (\ref{LL}) is simply equal to the total number of negation-configurations $\Phi(0)=K \alpha \log{2}$. For $x=1$, the partition function above is related to the annealed partition function, $\Phi(1)= \log{2}+ \alpha \log\left( 2^ K-1 \right)$.

The major difficulty in calculating $\Phi(x)$, for a general value of $x$, is that the entropy $s(\mathcal{J})$ is not defined as a sum of local terms. On the other hand the geometry of the underlying factor graph is still tree-like in the random AdSAT, hence for any configuration of negations $\mathcal{J}$ we can apply the cavity method (with replica symmetry breaking if needed) to compute the entropy $s(\mathcal{J})$. In the cavity method, as it is reminiscent of the Bethe approximation, the entropy (or more generally Bethe free energy) can be written as a sum of local terms. This fact enables us to calculate $\Phi(x)$. 

The statistical physics treatment of the random $K$-SAT problem among others led to a discovery that replica symmetry breaking approach is needed  \cite{MezardParisi01,MezardParisi02,KrzakalaMontanari06} in order to correctly compute the entropy close to the satisfiability threshold $\alpha_c$. Said in other words, in that region the space of solutions splits into well ergodically separated clusters. We define the complexity function, $\Sigma$, as the logarithm of the total number of clusters per variable. The value of the complexity can then be computed with the survey propagation equations \cite{MezardParisi02}. At the satisfiability threshold the complexity goes to zero, whereas the entropy density of solutions is a positive number even at the threshold. With this in mind it will be useful to define also 
\be
     e^{N\Phi_{SP}(x)} = \sum_{\mathcal{J}} e^{x N \Sigma(\mathcal{J})} = e^{N[{\cal L}_{SP}(\Sigma) + x\Sigma]}\, , \quad \quad  \frac{\partial {\cal L}_{SP}(\Sigma)}{\partial \Sigma} = -x \, .
\ee
where ${\cal L}_{SP}(\Sigma)$ is the entropy density of negation-configurations that give a certain complexity function $\Sigma$.

\section{Reminder of equations for belief and survey propagation}\label{sec_BP}

With the notation introduced in the previous Section we write the belief BP equations and the Bethe entropy as derived e.g. in \cite{YedidiaFreeman03,MezardMontanari07}. These equations are asymptotically exact on locally tree-like graphs as long as all correlation length-scales are finite. If they are not then splitting the phase space into clusters such that within each cluster the correlations decay again might be possible. SP then estimates the total number of such clusters \cite{MezardParisi02}, and it does so asymptotically exactly at least close enough to the satisfiability threshold \cite{MontanariParisi04,KrzakalaMontanari06}.

Denoting by $\{ m_{ia}, \hat{m}_{ai} \}$ the BP (SP) messages, we write the BP (SP) fixed point equations as 
\begin{eqnarray}
m_{ia} & = & g_{ia}(\{\hat{m}_{bi}\}_{b\in\partial i\setminus a},\{J_{bi}\}_{b\in\partial i}),\label{105}\\ \nonumber
\hat{m}_{ai} & = & \hat{g}_{ai}(\{m_{ja}\}_{j\in\partial a\setminus i},\{J_{ja}\}_{j\in\partial a}).
\end{eqnarray}
In the BP case, the messages read \cite{MezardMontanari07} $m_{ia}=\{\nu_{ia}^0,\nu_{ia}^1\}$, $\hat{m}_{ai}=\{\hat{\nu}_{ai}^0,\hat{\nu}_{ai}^1\}$, and 
\begin{eqnarray}
g^r_{ia}(\{\hat{\nu}_{bi}\}_{b\in\partial i\setminus a}) & = & \frac{\prod_{b\in\partial i\setminus a}\hat{\nu}_{bi}^r}{\prod_{b\in\partial i\setminus a}\hat{\nu}_{bi}^0+\prod_{b\in\partial i\setminus a}\hat{\nu}_{bi}^1},\label{BP_eq1}\\ \nonumber
\hat{g}^r_{ai}(\{\nu_{ja}\}_{j\in\partial a\setminus i},\{J_{ja}\}_{j\in\partial a}) & = & \frac{1-\delta_{r,J_{ai}}\prod_{j\in\partial a\setminus i}\nu_{ja}^{J_{ja}}}{2-\prod_{j\in\partial a\setminus i}\nu_{ja}^{J_{ja}}}, \label{BP_eq2}
\end{eqnarray}
where $r=0,1$.
In the SP case, $m_{ia}=\{Q_{ia}^ {S}, Q_{ia}^ U, Q_{ia}^\ast\}, \, \hat{m}_{ai}=\hat{Q}_{ai}$ and 
\begin{eqnarray}\label{85}
g_{ia}^{\ast}(\{J_{bi}\}_{b\in\partial i},\{\hat{Q}_{bi}\}_{b\in\partial i\setminus a}) & = & C \prod_{b \in \partial i \setminus  a} (1-\hat{Q}_{bi}),\\ \nonumber
g_{ia}^{S}(\{J_{bi}\}_{b\in\partial i},\{\hat{Q}_{bi}\}_{b\in\partial i\setminus a}) & = & C\prod_{b \in \mathcal{U}_{ia}} (1-\hat{Q}_{bi})\left[ 1- \prod_{b \in \mathcal{S}_{ia}}(1-\hat{Q}_{bi})\right],\\ \nonumber
 g_{ia}^{U}(\{J_{bi}\}_{b\in\partial i},\{\hat{Q}_{bi}\}_{b\in\partial i\setminus a}) & = & C \prod_{b \in \mathcal{S}_{ia}} (1-\hat{Q}_{bi})\left[ 1- \prod_{b \in \mathcal{U}_{ia}}(1-\hat{Q}_{bi})\right]\\ \nonumber
\hat{g}_{ai}(\{Q_{ja}\}_{j\in\partial a\setminus i}) & = & \prod_{j\in \partial a \setminus  i} Q_{ja}^{U},
 \end{eqnarray}
where $C$ is a normalization constant enforcing the relation $g_{ia}^{\ast} + g_{ia}^{S} +g_{ia}^{U}=1$, and $\mathcal{S}_{ia},\, \mathcal{U}_{ia}$ are defined as
\be
\begin{array}{ccc}
\text{if}\ J_{ia}=0   & \mathcal{S}_{ia}= \partial_0 i \textbackslash \{ a \}, \, \mathcal{U}_{ia} = \partial_1 i\\
\text{if}\ J_{ia}=1  & \mathcal{S}_{ia}= \partial_1 i \textbackslash \{ a \}, \, \mathcal{U}_{ia} = \partial_0 i,
\end{array}
\ee
where $\partial_{0/1}i=\{ a \in \partial i \, \text{such that}\, J_{ia}=0/1 \}$.

If $\{m_{ia},\hat{m}_{ai} \}$ is a fixed point of Eqs.~(\ref{105}), the Bethe entropy for BP and the complexity for SP are both be written in a general form 
\be 
s(\{J_{ia},m_{ia}, \hat{m}_{ai} \}) =\sum_{a=1}^{M}\mathbb{S}_a(\{m_{ia},J_{ia}\}_{i\in\partial a}  ) + \sum_{i=1}^{N}\mathbb{S}_i(\{\hat{m}_{ai},J_{ia}\}_{a\in\partial i}  ) - \sum_{(a,i)}\mathbb{S}_{ai}(m_{ia},\hat{m}_{ai}) \label{entropy}
\ee
where for BP
\bea
      \mathbb{S}_a(\{\nu_{ia},J_{ia}\}_{i\in\partial a}  ) &= &\log \left( 
1- \prod_{i \in \partial a} \nu_{ia}^{J_{ia}} \right),\nonumber \\
      \mathbb{S}_i(\{\hat{\nu}_{ai},J_{ia}\}_{a\in\partial i}  ) &=&   \log \left(   \prod_{b \in \partial i} \hat{\nu}_{bi}^0 + \prod_{b \in \partial i} \hat{\nu}_{bi}^1  \right),\\
      \mathbb{S}_{ai}(\nu_{ia},\hat{\nu}_{ai}) &=& \log \left( \nu_{ia}^0 \hat{\nu}_{ai}^0 + \nu_{ia}^1 \hat{\nu}_{ai}^1 \right) ,\nonumber
\eea
while for SP
\bea
      \mathbb{S}_a(\{Q_{ia},J_{ia}\}_{i\in\partial a}  )& =& \log \left( 1 - \prod_{j \in \partial a} Q_{ja}^ U \right), \nonumber \\
      \mathbb{S}_i(\{\hat{Q}_{
      ai},J_{ia}\}_{a\in\partial i}  )& =& \log \left[ \prod_{ b \in \partial _0 i} (1 - \hat{Q}_{bi} ) +\prod_{ b \in \partial _1 i} (1 - \hat{Q}_{bi} )-\prod_{ b \in \partial i} (1 - \hat{Q}_{bi} ) \right]  ,  \\
      \mathbb{S}_{ai}(Q_{ia},\hat{Q}_{ai}) &= & \log \left( 1 - Q_{ia}^ U \hat{Q}_{ai} \right).\nonumber
\eea

\section{Computation of  the large deviations function}\label{sec_deriv}

The most important formula of the previous Section is (\ref{entropy}), in certain regimes it gives the asymptotically exact entropy or complexity in a form factorized in local terms. The remaining complication is that now everything depends on the fixed point of the BP (SP) equations. We can, however, write
\bea
Z(x) &= & \sum_{\mathcal{J}} \int \prod_{ia} dm_{ia} d\hat{m}_{ai} e^{Nx  s(\{J_{ia},m_{ia}, \hat{m}_{ai} \})} \times\\
&& \times \prod_{(i,a)} \delta(m_{ia} - g_{ia}(\{\hat{m}_{bi}\}_{b\in\partial i\setminus a},\{J_{bi}\}_{b\in\partial i}))
\prod_{(a,i)}  \delta( \hat{m}_{ai} - \hat{g}_{ai}(\{m_{ja}\}_{j\in\partial a\setminus i},\{J_{ja}\}_{j\in\partial a})),\nonumber
\eea
If we now introduce auxiliary variables $\omega_{ia}\equiv\{J_{ia},m_{ia}, \hat{m}_{ai} \}$ the free energy defined by (\ref{LL}) can be re-written in the common local form
\begin{eqnarray} \label{1000}
Z(x) = \sum_{\{\omega_{ia}\}} \left\{ \left[\prod_{a=1}^{M}\Psi_{a}(\{\omega_{ia}\}_{i\text{\ensuremath{\in\partial a}}})\right]\left[\prod_{i=1}^{N}\Psi_{i}(\{\omega_{ia}\}_{a\in\partial i})\right]\left[\prod_{(a,i)}\Psi_{ai}(\omega_{ia})\right] \right\}  .  
\end{eqnarray}
where
\begin{eqnarray} \nonumber
\Psi_a(\{\omega_{ia}\}_{i\in \partial a} ) & \equiv & e^{x\mathbb{S}_a(\{m_{ia},J_{ia}\}_{i\in\partial a}  )} 
\prod_{i\in \partial a} \delta (\hat{m}_{ai}  - \hat{g}_{ai}(\{m_{ja}\}_{j\in\partial a\setminus i},\{J_{ja}\}_{j\in\partial a})),\\ \label{108} 
\Psi_i(\{\omega_{ia}\}_{a\in \partial i} ) & \equiv & e^{x\mathbb{S}_i(\{\hat{m}_{ai},J_{ai}\}_{a\in\partial i}  )} 
\prod_{a\in \partial i} \delta(m_{ia}  -  g_{ia}(\{\hat{m}_{bi}\}_{b\in\partial i\setminus a},\{J_{bi}\}_{b\in\partial i})),\\ \nonumber
\Psi_{ai}(\omega_{ia}) & \equiv & e^{-x\mathbb{S}_{ai}(m_{ia},\hat{m}_{ai})}.
\end{eqnarray}
In Eq. (\ref{1000}) and in the following, the sum over $\omega_{ia}$ stands for the sum over $J_{ia}$ and the integral over $m_{ia}$, $\hat{m}_{ai}$. 

\begin{figure} 
\centering  
\includegraphics[scale=0.8]{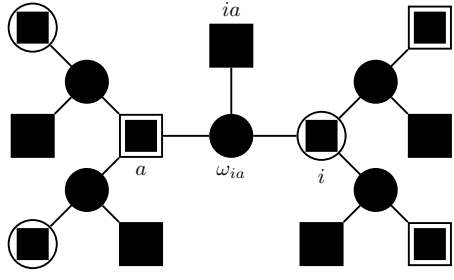}
\caption{The auxiliary factor graph describing Eq. (\ref{1000}). Empty  circles and empty squares  represent variable and function nodes of the original factor graph of random $K$-SAT.
Empty squares  with black-filled squares inside represent the $\Psi_{a}$ function nodes, empty circles with  black-filled squares inside represent $\Psi_{i}$ function nodes of the auxiliary graphs. Finally, black-filled squares represent $\Psi_{ia}$ function nodes, while black-filled circles $\omega_{ia}$-variable nodes of the auxiliary graph.  An edge connecting $\omega_{ia}$ to a function node of the auxiliary graph means that such a function node depends on $\omega_{ia}$.}   
\label{fig2}  
\end{figure}

The probability measure in Eq. (\ref{1000}) is local and can be hence represented with an auxiliary factor graph that can be viewed as decorating the original $K$-SAT factor graph. Fig. \ref{fig2} depicts this construction. 

The partition function $Z(x)$ can now be computed by implementing the general BP formalism to the auxiliary graph, just as it is done in the derivation of the 1RSB equations in \cite{MezardMontanari07} (note indeed the close formal resemblance of our approach and the 1RSB equations). 
We call $S_{ia}(\omega_{ia})$ the message going from the variable node ${ia}$ to the function node $a$, and $\hat{S}_{ai}(\omega_{ia})$ 
the message going from the variable node ${ia}$ to the function node
$i$. BP equations on the auxiliary factor graph on the variables $\omega_{ia}$ then lead to fixed-point equations for these messages
\begin{eqnarray}  \nonumber
\hat{S}_{ai}(\omega_{ia}) & \simeq & \sum_{\{\omega_{ja}\}_{j\in\partial a\setminus i}}[\hat{z}_{ai}(\{m_{ja},J_{ja}\}_{j\in\partial a},\hat{m}_{ai})]^{x}\prod_{j\in\partial a}I(\hat{m}_{aj}=\hat{g}_{aj}(\{m_{ka}\}_{k\in\partial a\setminus j},\{J_{ka}\}_{k\in\partial a}))\times \\ \label{113}
&&\times  \prod_{j\in\partial a\setminus i}S_{ja}(\omega_{ja}),\\ \nonumber
S_{ia}(\omega_{ia}) & \simeq & \sum_{\{\omega_{ib}\}_{b\in\partial i\setminus a}}[z_{ia}(\{\hat{m}_{bi},J_{ib}\}_{b\in\partial i},m_{ia})]^{x}\prod_{b\in\partial i}I(m_{ib}=g_{ib}(\{\hat{m}_{ci}\}_{c\in\partial i\setminus b}, \{J_{ci}\}_{c\in\partial i} )) \times \\ \nonumber
&& \times  \prod_{b\in\partial i\setminus a}\hat{S}_{bi}(\omega_{ib}),\nonumber
 \end{eqnarray}
where
\begin{eqnarray*}
z_{ia}(\{\hat{m}_{bi},J_{ib}\}_{b\in\partial i},m_{ia}) & \equiv & e^{\mathbb{S}_{i}(\{\hat{m}_{bi},J_{ib}\}_{b\in\partial i})-\mathbb{S}_{ai}(m_{ia},\hat{m}_{ai})},\\
\hat{z}_{ai}(\{m_{ja},J_{ja}\}_{j\in\partial a},\hat{m}_{ai}) & \equiv & e^{\mathbb{S}_{a}(\{m_{ia},J_{ia}\}_{i\in\partial a})-\mathbb{S}_{ai}(m_{ia},\hat{m}_{ai})}.
\end{eqnarray*}

Eq.~(\ref{113}) can be further simplified. It is easy to check that when the fixed point Eqs. (\ref{105}) hold, the term $z_{ia}$ (resp. $\hat{z}_{ai}$) does not depend on the `backward' messages $m_{ia}$ (resp. $\hat{m}_{ai}$). Using this result, we can see that (\ref{113})
are compatible with a choice of the messages $S_{ia},\,\hat{S}_{ai}$
depending \textit{only} on $m_{ia},J_{ia}$ and only on $\hat{m}_{ai},J_{ia}$
respectively. Indeed, if we assume that $S_{ia}(\omega_{ia}) = S_{ia}(m_{ia},J_{ai})$, and $\hat{S}_{ai}(\omega_{ia}) = \hat{S}_{ai}(\hat{m}_{ai},J_{ai})$ then (\ref{113}) becomes
\begin{eqnarray}
S_{ia}(m_{ia},J_{ai})  &\simeq  & \sum_{\{J_{bi}\}_{b\in\partial i\setminus a}} \int \prod_{b\in\partial i\setminus a} d\hat{m}_{bi} \hat{S}_{bi}(\hat{m}_{bi},J_{bi})   \,  [z_{ia}(\{\hat{m}_{bi},J_{ib}\}_{b\in\partial i\setminus a})]^{x} \times \label{10} \\ \nonumber
&& \times  \delta(m_{ia}- g_{ia}(\{\hat{m}_{bi}\}_{b\in\partial i\setminus a},\{J_{bi}\}_{b\in\partial i}))\\
\hat{S}_{ai}(\hat{m}_{ai},J_{ai}) & \simeq & \sum_{\{J_{aj}\}_{j\in\partial a\setminus i}} \int  \prod_{j\in\partial a\setminus i} dm_{ja} S_{ja}(m_{ja},J_{ja})  \, [\hat{z}_{ai}(\{m_{ja},J_{ja}\}_{j\in\partial a\setminus i})]^{x}   \times  \label{11} \\  \nonumber
&& \times \delta(\hat{m}_{ai}-\hat{g}_{ai}(\{m_{ja}\}_{j\in\partial a\setminus i},\{J_{ja}\}_{j\in\partial a})).
\end{eqnarray}

The free energy $\Phi(x)$ can then be computed using the general expression
for the Bethe free-entropy \cite{MezardMontanari07}
\begin{equation}\label{17}
N\Phi(x)=\sum_{a=1}^{M}\mathbb{F}_{a}+\sum_{i=1}^{N}\mathbb{F}_{i} -\sum_{(i,a)}\mathbb{F}_{ia},
\end{equation}
where
\begin{eqnarray}\label{18}
\mathbb{F}_{a} & = & \log\left[\sum_{\{J_{ia}\}_{i\in\partial a}} \int  \prod_{i\in\partial a} dm_{ia}  S_{ia}(m_{ia},J_{ia}) e^{x\mathbb{S}_{a}(\{m_{ia},J_{ia}\}_{i\in\partial a})}\right],\\
\label{19}
\mathbb{F}_{i} & = & \log\left[\sum_{\{ J_{ia}\}_{a\in\partial i}} \int  \prod_{a \in \partial i} d\hat{m}_{ai}  \hat{S}_{ai}(\text{\ensuremath{\hat{m}}}_{ai},J_{ia}) e^{x\mathbb{S}_{i}(\{\hat{m}_{ai}\}_{a\in\partial i})} \right],\\
\mathbb{F}_{ia} & = & \log\left[\sum_{J_{ia}} \int d m_{ia} d \hat{m}_{ai} S_{ia}(m_{ia},J_{ia})\hat{S}_{ai}(\text{\ensuremath{\hat{m}}}_{ai},J_{ia}) e^{x\mathbb{S}_{ia}(m_{ia},\hat{m}_{ai})}\right].\label{20}
\end{eqnarray}

Equations (\ref{10}-\ref{11}) clearly show the formal analogy  with $1$RSB cavity equations \cite{MezardMontanari07}. The difference between  Eqs. (\ref{10}-\ref{11}) and the latter is that in the AdSAT case the negations are considered as physical degrees of freedom of the partition function $Z(x)$, and the resulting cavity equations (\ref{10}-\ref{11}) consist in a weighted average of the quantities  $[z_{ia}(\{\hat{m}_{bi},J_{ib}\}_{b\in\partial i\setminus a})]^{x}$, $[\hat{z}_{ai}(\{m_{ja},J_{ja}\}_{j\in\partial a\setminus i})]^{x}$ over $\{J_{ia},m_{ia}, \hat{m}_{ai} \}$. On the contrary, in the $1$RSB case the only degrees of freedom are the BP messages $\{m_{ia}, \hat{m}_{ai} \}$ in such a way that the resulting 1RSB cavity equations consist in an average over $m_{ia}, \hat{m}_{ai}$ at fixed $J_{ia}$s.  

The biggest advantage of the formal resemblance of Eqs. (\ref{10}-\ref{11}) to the 1RSB cavity equations is that in order to solve (\ref{10}-\ref{11}) numerically we can use the very same technique and all the related knowledge as in the case of 1RSB.  
We indeed use the population dynamics \cite{MezardParisi01}, where the distributions $S_{ia}(m_{ia},J_{ia}),\, \hat{S}_{ai}(\hat{m}_{ai}, J_{ia})$ are represented as populations of $P$ messages.
When the size of the population $P$ is large, we expect the populations to reproduce well the distributions $S_{ia}(m_{ia}, J_{ia})$, and  $\hat{S}_{ai}(\hat{m}_{ai}, J_{ia})$. The cavity equations (\ref{10}-\ref{11}) can be written in terms of these populations. Starting from a given initial configuration, the iteration of the cavity equations yields the fixed-point populations satisfying (\ref{10}-\ref{11}). Once this fixed point is achieved the free-energy $\Phi(x)$ can be computed numerically by means of (\ref{17}-\ref{20}). This is repeated for different values of $x$ and finally the Legendre transform ${\cal L}(s)$ is evaluated.
Everything is done in the very same way the 1RSB equations are usually solved, for more details see e.g. \cite{MezardParisi01,ZdeborovaKrzakala07,MontanariRicci08}. The only difference is in the treatment of the negation-variables. 
In our case one  writes the distributions $S_{ia}(m_{ia},J_{ia}), \hat{S}_{ai}(\hat{m}_{ai},J_{ia}) $ as 
\[
S_{ia}(m_{ia},J_{ia}) = \frac{1}{2} S_{ia}(m_{ia} | J_{ia}), \, \quad \quad \hat{S}_{ai}(\hat{m}_{ai},J_{ia}) = \frac{1}{2} \hat{S}_{ai}(\hat{m}_{ai}| J_{ia}),
\]
because $S_{ia}(J_{ia}) = \int {\rm d}m_{ia} S_{ia}(m_{ia},J_{ia}) = 1/2=  \hat{S}_{ai}(J_{ia}) $ as can be seen by explicitly integrating Eqs.~(\ref{10}-\ref{11}). 
We then introduce a pair of populations  $\{ S_{ia}^{J}[s], \hat{S}_{ai}^{J}[s] \}_{s=1,\cdots,P}$ representing the probability distributions $S_{ia}(m_{ia} | J), \, \hat{S}_{ai}(\hat{m}_{ai}| J)$ respectively. The population dynamics is then implemented in terms of such populations, and the resulting fixed point investigated numerically. 

\section{Cavity method results for random AdSAT}\label{sec_res}

In this Section we present the solution of the cavity equations (\ref{10}-\ref{11}), and its implications for the random AdSAT problem. 

\subsection{Large deviations of the entropy and complexity on regular instances}\label{sub_reg}

Before addressing the random AdSAT as defined in Sec.~\ref{sec_intro} we will study it on random regular instances. On $L$-regular instances every variable belongs to exactly $L$ clauses. A random $L$-regular instance is chosen uniformly at random from all possible ones with given number of variables $N$ and number of clauses $M$, provided that $KM=LN$.

In the $3$-SAT problem discussed in this paper the Bethe entropy is asymptotically exact only as long as the BP equations converge to a fixed point \cite{KrzakalaMontanari06}, the non-convergence is equivalent to the spin glass instability, or a continuous transition to replica symmetry breaking phase. On random regular graphs with random values of negations, BP stops converging at $L=12$ (this is the reason why these and larger value are omitted from Table~\ref{table_reg}), meaning that for $L\ge 12$ there is a need for SP (or other form of replica symmetry breaking solution). For $L\ge 15$ BP iterations for random regular 3-SAT lead to contradictions (zero normalizations) meaning that in this region the large random instances are almost surely unsatisfiable.  Survey propagation on random regular 3-SAT has a trivial fixed point for $L\le 12$, a fixed point with for $L=13$ with the value of complexity $\Sigma(L=13)=0.008$. SP does not converge for $L\ge 14$, if we ignore non-convergence, and compute the complexity from the current values of  messages we get on average $\Sigma(L=14)=-0.03$. This means that $L=13$ is the largest satisfiable case. 

The great advantage of random regular instances is that topologically the local neighborhood looks the same for every variable~$i$. Moreover, we remind that in regimes where the BP equations are asymptotically exact properties of variable~$i$ depend only on the structure of the local neighborhood of $i$. 
Hence on regular graphs all the quantities in Eqs.~(\ref{10}-\ref{11}) are independent of the indices $i,j$, and $a,b$. This so-called {\it factorization} property simplifies crucial numerical solution of Eqs. (\ref{10}-\ref{11}), the $2 K M$  distributions  $\{ S_{ia}(\nu_{Ia},J_{ia}), \hat{S}_{ai}(\hat{\nu}_{ai}, J_{ai} ) \}$ reduce to only $2$ distributions $ S(\nu,J), \, \hat{S}(\hat{\nu}, J )$. Moreover, the thermodynamic limit is taken directly without increasing the computational effort. (To avoid confusion, we remind here that for the canonical random $K$-SAT problem where negation-variables are chosen uniformly at random and fixed, the BP solution is not factorized. In the adversarial version one sums over the negation-variables, hence the factorization.)

\begin{figure}
\centering
\includegraphics[scale=0.6]{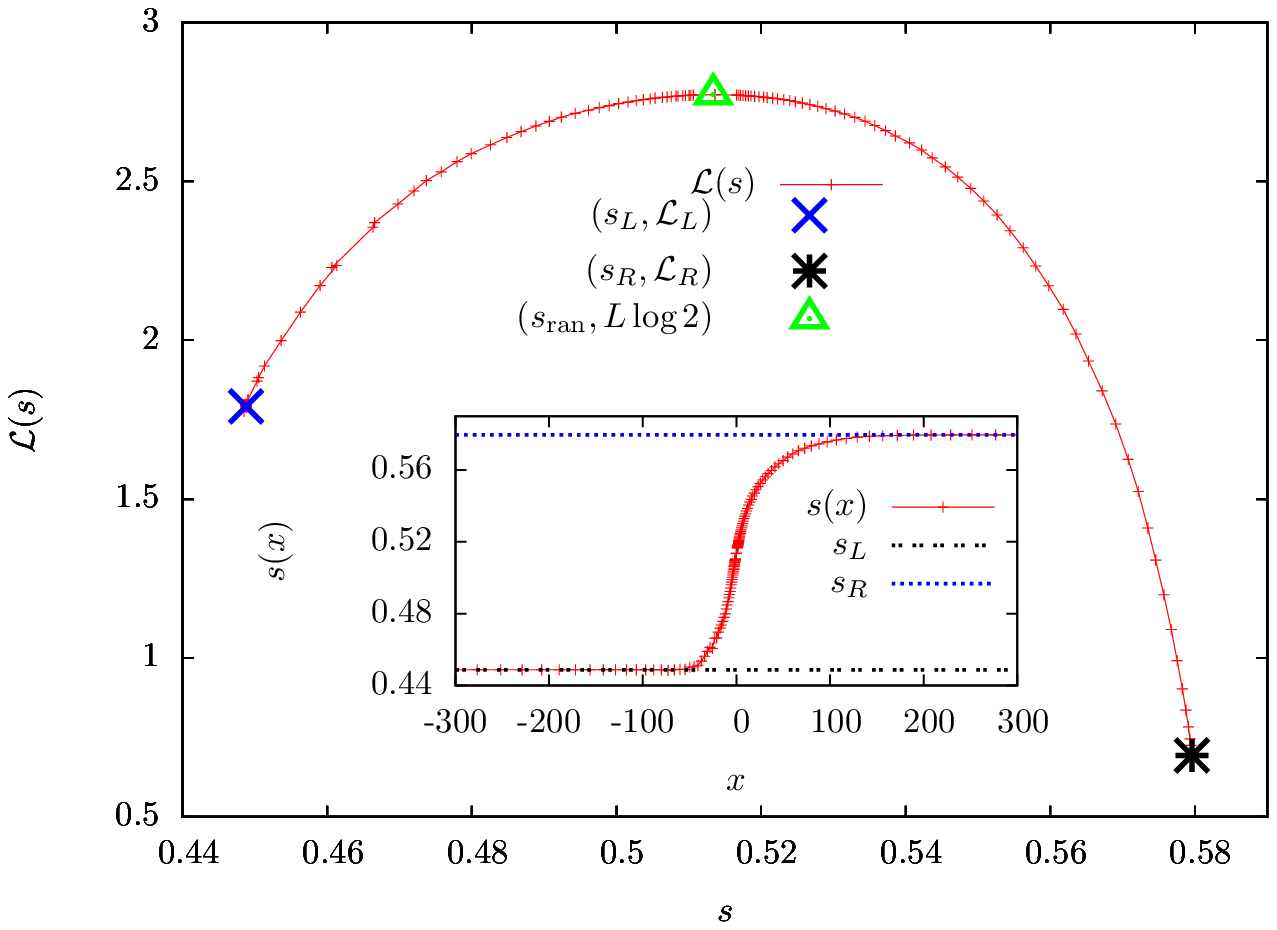}
\includegraphics[scale=0.6]{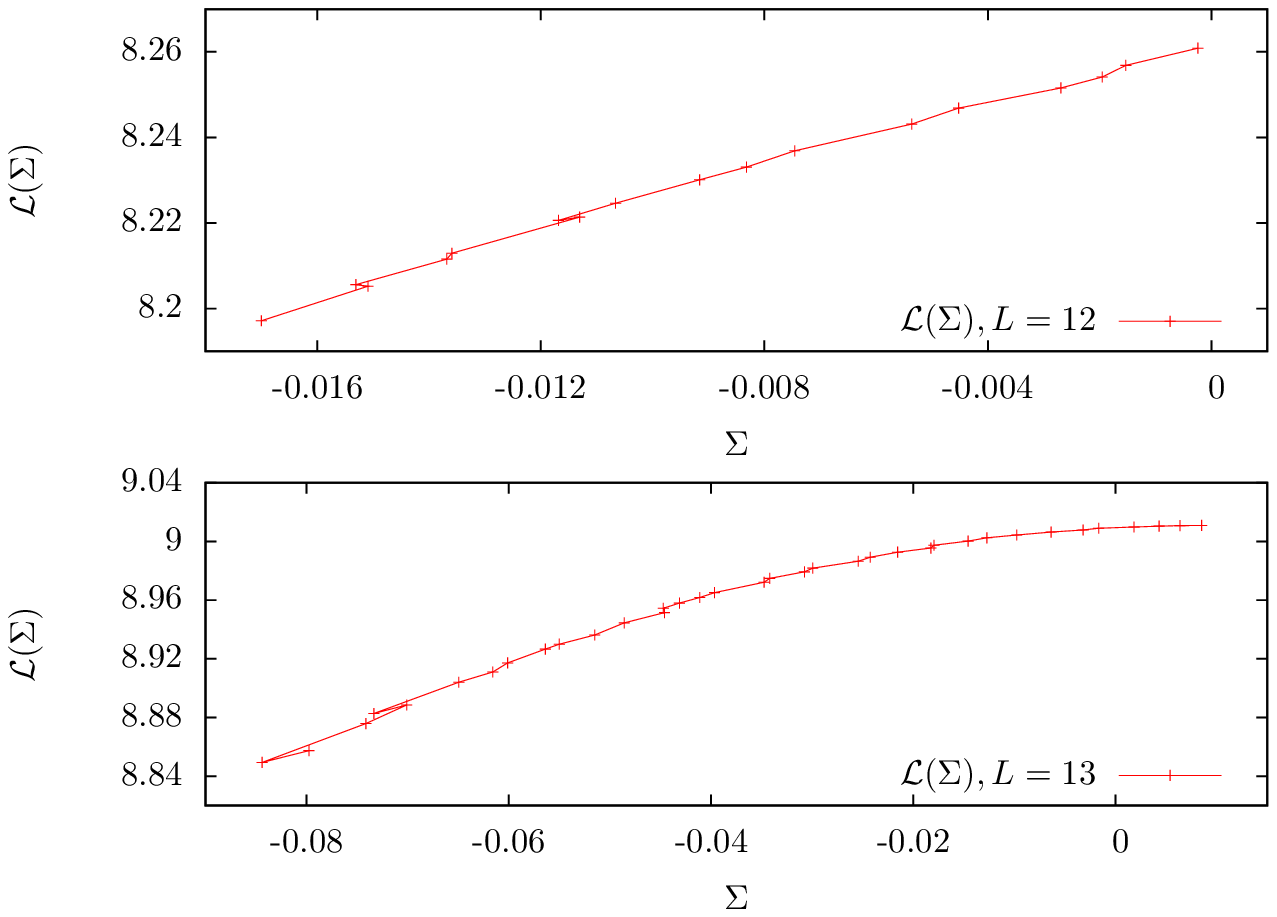}
\caption{Left: The large deviation function $\mathcal{L}(s)$ vs. Bethe entropy $s$ computed by population dynamics on regular graphs with $K=3$, $L=4$, population size $P=10^4$.  The left ending point $(f_L, \mathcal{L}_L)$ corresponds to balanced configurations of negations, whereas the right ending point $(f_R, \mathcal{L}_R)$ to the polarized configurations of negations (more in the text). 
Right: The large deviation function $\mathcal{L}_{SP}(\Sigma)$ vs. the complexity $\Sigma$ computed by population dynamics on regular graphs with $K=3,\, L=13,\, P=10^4$ (bottom), and $K=3,\, L=12,\, P=7.5 \cdot 10^4$ (top). }
\label{fig08}
\end{figure}

In Fig.~\ref{fig08} we show the large deviation function $\mathcal{L}(s)$ of the Bethe entropy $s$ obtained by the population dynamics over BP messages on regular graphs with $K=3$ and variable degree $L=4$, i.e. by solving Eqs.~(\ref{10}-\ref{11}) and (\ref{17}-\ref{20}). First of all, in the ``infinite temperature'' case, i.e. when the Legendre parameter $x=0$, that is at the maximum of $\mathcal{L}(s)$, we recover the logarithm of the total number of negation-configurations $\mathcal{L}(x=0)= L \log{2}$. The corresponding value of entropy  $s(x=0)=s_{\rm ran}$ is the Bethe entropy for a random choice of negations (values summarized in Table~\ref{table_reg}).

\begin{table}
\centering
\begin{tabular}{|c|c|c|c|}
\hline 
$L$ & $s_{\rm ran}$ & $s_{B}$ & $s_{R}$\tabularnewline
\hline
\hline 
$2$ & $0.6039$ & $0.5710$ & 0.6196\tabularnewline
\hline 
$3$ & $0.5592$ & $0.5324$ & 0.5975\tabularnewline
\hline 
$4$ & $0.5134$ & $0.4488$ & 0.5796\tabularnewline
\hline 
$5$ & $0.4686$ & $0.4120$ & 0.5644\tabularnewline
\hline 
$6$ & $0.4220$ & $0.3266$ & 0.5513\tabularnewline
\hline 
$7$ & $0.3750$ & $0.2902$ & 0.5397\tabularnewline
\hline 
$8$ & $0.3302$ & $0.2044$ & 0.5293\tabularnewline
\hline 
$9$ & $0.2816$ & $0.1677$ & 0.5199\tabularnewline
\hline 
$10$ & $0.2319$ & $0.082$ * & 0.5114\tabularnewline
\hline 
$11$ & $0.1813$ & $0.042$ * & 0.5035 \tabularnewline
\hline
$12$ & $0.128$ * & $\times$ & $0.4962$ \tabularnewline
\hline
$13$ & $0.07$ * & $\times$ & $0.4894$ \tabularnewline
\hline
$14$ & $\times$ & $\times$ & $0.4831$ \tabularnewline
\hline
\end{tabular}
 
\caption{
\label{table_reg} Bethe free-entropy on regular instances with random negation-configuration ($s_{\rm{ran}}$),  with balanced negation-configurations ($s_B$), and non-frustrated with $J_{ia}=0 \, \forall i,a$ ($s_{R}$). The entropy for the non-frustrated case and for the balanced case for even degree $L\le 10$ can be computed analytically since the BP fixed is factorized in these cases. In the other cases we iterate BP equations on large random graphs and compute the entropy from the corresponding BP fixed point. The star signals that BP did not converge and the value of entropy was obtained by averaging over an interval of time. The $\times$ means that BP converged to contradictions for these densities of constraints.}
\end{table}

The inset of the figure shows that as the Legendre parameter $x\to \pm \infty$ both $\mathcal{L}$ and $s$ converge to well defined ending points (the same data in a logarithmic plot show that the convergence is exponential). Let us denote the lowest entropy ending point (left, $x\to -\infty$) $(s_L,\mathcal{L}_L)$, and the highest entropy ending point (right, $x\to \infty$) $(s_R,\mathcal{L}_R)$. We observe systematically that the value of $s_R=s_U$, where $s_U$ is the entropy of the uniform negation-configuration which is obtained by computing a fixed point of Eqs. (\ref{BP_eq1}-\ref{BP_eq2}) such that $\hat{\nu}_{bi}=\hat{\nu}\,\,  \forall b,i$, ${\nu}_{ja}={\nu}\, \, \forall j,a$ and $J_{ia}=0\,\, \forall i,a$, and plugging it into Eq.~(\ref{entropy}), values summarized in Table~\ref{table_reg}. An edge independent fixed point of the BP equations is called factorized. We realize that $s_U$ also corresponds to the value of the Bethe entropy when every variable is either always negated or never negated, we call such negation-configurations {\it polarized}. There are $2^{LN}$ polarized negation-configurations, and indeed the logarithm of the number of such choices corresponds to the value of $\mathcal{L}_R = \log{2}$. Intuitively such configurations of negation are frustrating the formula in the least possible way. And Fig.~\ref{fig08} shows that such intuition is asymptotically exact in this case. 

Similarly for the lowest entropy ending point, for even values of the degree $L$, we realize that $\mathcal{L}_L=\log{L\choose L/2}$ and $s_L$ corresponds to a value $s_B$ that is obtained from a factorized BP fixed point when each variable is $L/2$ times negated and $L/2$ times non-negated, values are summarized in Table~\ref{table_reg}. Such {\it balanced} configurations of negations locally frustrate the variables in a maximal way (half clauses want the variable to be $1$, the other half $0$). And the computation presented in Fig.~\ref{fig08} suggests that asymptotically there are no correlated negation configurations that would frustrate the formula even more and decrease the value of the entropy further. 

We investigate more in detail the result following from Fig.~\ref{fig08}, i.e. that the most frustrated configurations of negations on the regular graphs with even degree are the balanced negations, we denote the balanced negation-configurations $\{J_{ia}\}_B={\mathcal J}_B$. There is ${L\choose L/2}^N$ of such negation-configurations. Does our result mean that all of them lead to the same number of solutions ${\cal N}(\mathcal{J}_B)$? We will see in Sec.~\ref{sec_num} that this is not true  for finite $N$. The correct conclusion of the result presented in Fig.~\ref{fig08} is that $\lim_{N\to \infty} [\log{{\cal N}(\mathcal{J}_B)}]/N = s_B = s_L$ independently of the realization of ${\mathcal J}_B$. This can also be seen directly from the solution of the BP equations on the formulas with balanced negations. Indeed, the fixed point of the BP equations is factorized and independent of the realization of negations and also of the size of the graph. We tried numerically formulas of various sizes and many possible realizations of balanced negations and for even degree $L<10$ BP always converges to the factorized fixed point (at $L=10$ BP stops converging, as we will discuss later in the paper) giving always the same Bethe entropy density $s_B$. Further discussion about the true entropy fluctuations compared to the constant Bethe entropy in this case will be presented in Sec.~\ref{sec_num}    

For regular graphs with odd degree $L$ we cannot achieve ideal balancing of every variable. Instead, we call a configuration of negations balanced if for every variables there is either $(L-1)/2$ or $(L+1)/2$ negations. The total number of such configurations is then $2^N{L\choose (L-1)/2}^N$. BP fixed point on the balanced instances for odd $L$ is not factorized anymore. We can, however, solve the cavity equations (\ref{10}-\ref{11}) restricted to only balanced values of negations and we obtain that within the error-bars of the numerical resolution of the equations (that are less than 1\%) all the balanced configurations give the same value of Bethe entropy also in the odd $L$ case.

Our results for large deviations of the Bethe entropy lead to a conclusion that for the regular instances and in the limit $N\to \infty$ the most frustrated formulas are all those with balanced configurations of negations. Let us hence conclude this Section by summarizing the properties of regular SAT instances with balanced negations. BP on balanced instances converges for $L\le 9$, and leads to contradictions for $L\ge 13$. Survey propagation on balanced regular instances has a trivial fixed point for $L\le 9$, for $L=10$ a fixed point with complexity $\Sigma_B(L=10)=0.018$, for $L\ge 11$ the complexity is negative (e.g. $\Sigma_B(L=11)=-0.001$, $\Sigma_B(L=12)=-0.075$). 

For completeness, we also computed the large deviations of the complexity function. That is, we solved Eqs.~(\ref{10}-\ref{11}) using SP as the basic message passing scheme. Fig.~\ref{fig08} right shows some of the results for $L=12$ and $L=13$, we indeed see that there are configurations of negations that lead to negative complexity. Unfortunately, it is hard to extract any information from these curves for very negative values of $x$, because of the noise introduced by the finite population-size effects. This also poses a problem for $L=10$ and $11$ where we know that a non-trivial fixed point of SP exists for the balanced configurations of negations. In the population dynamics we should hence see a non-trivial solution for very negative values of $x$. Instead we were only able to obtain very noisy and inconclusive data from the population dynamics with population sizes up to $7.5\cdot 10^4$. For $L=10$ the SP equations have only one factorized fixed point for all the balanced configurations of negations, this again strongly suggests that instances with balanced negations are the most frustrated ones, and hence that for $L=10$ the adversary cannot make large formulas unsatisfiable. For lower values of $L\le 9$ the population dynamics has always only a trivial fixed point given by $S_{ia}(Q_{ia},J_{ia})= \delta(Q_{ia}^ S)\delta (Q_{ia}^ U)\delta ( Q_{ia}^ \ast-1)/2$, $\hat{S}_{ai}(\hat{Q}_{ai},J_{ia})=\delta(\hat{Q}_{ai})/2$ yielding $\Phi(x)=K \alpha \log 2$. SP is hence not very useful in this case to obtain new information about the random AdSAT problem.

In summary, for $L\ge 11$ the adversary will succeed to make a large formula unsatisfiable by simply balancing the negations (for $L\ge 14$ a random choice of negation would do). On the other hand, following our previous conclusion that the balanced formulas are the most frustrated ones, for $L\le 10$ the adversary will not be able to make large random regular SAT instances unsatisfiable by adjusting the values of negation-variables. 

\subsection{Results for random AdSAT, i.e. instances with Poisson degree distribution}

\begin{figure} 
\centering
\includegraphics[scale=0.6]{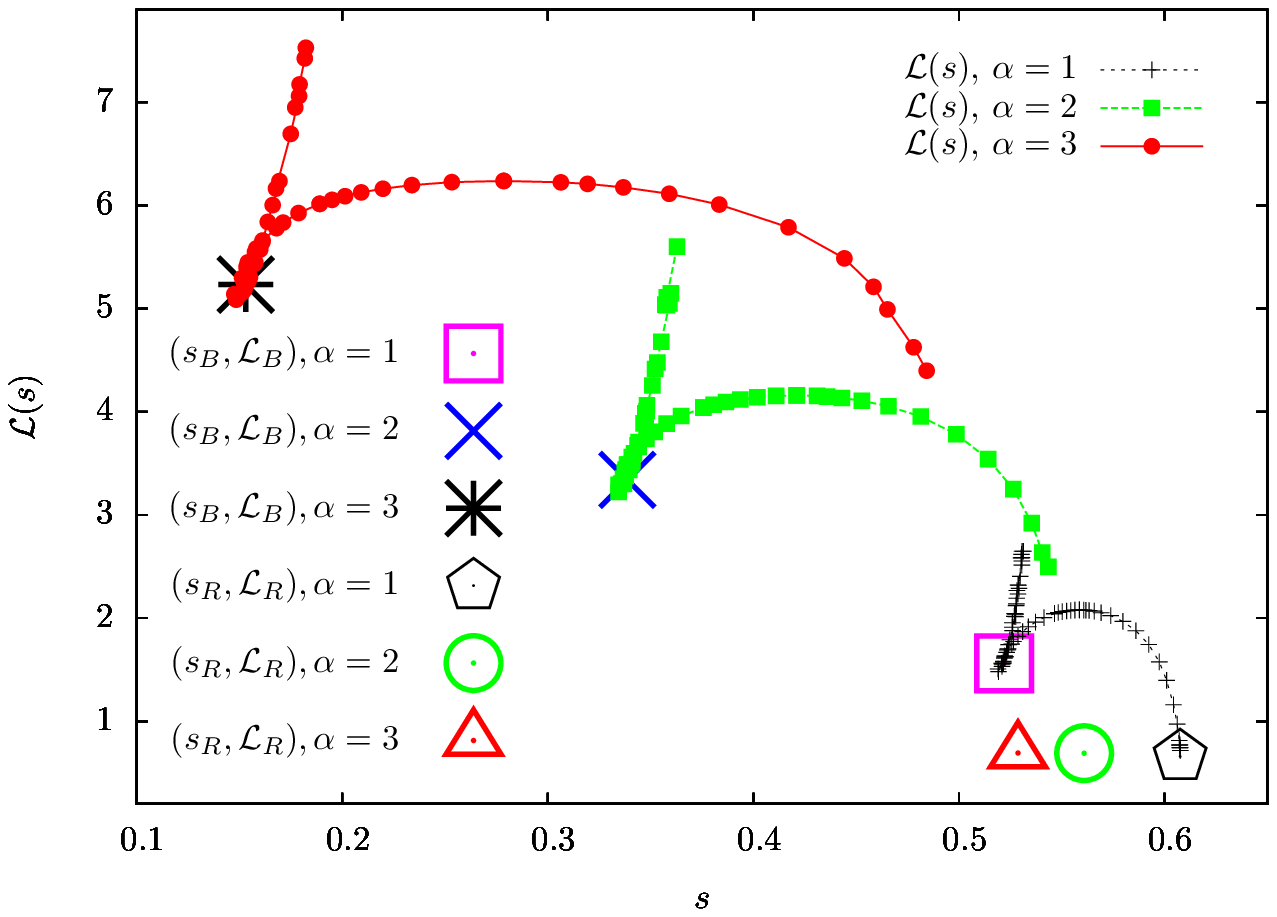}
\includegraphics[scale=0.6]{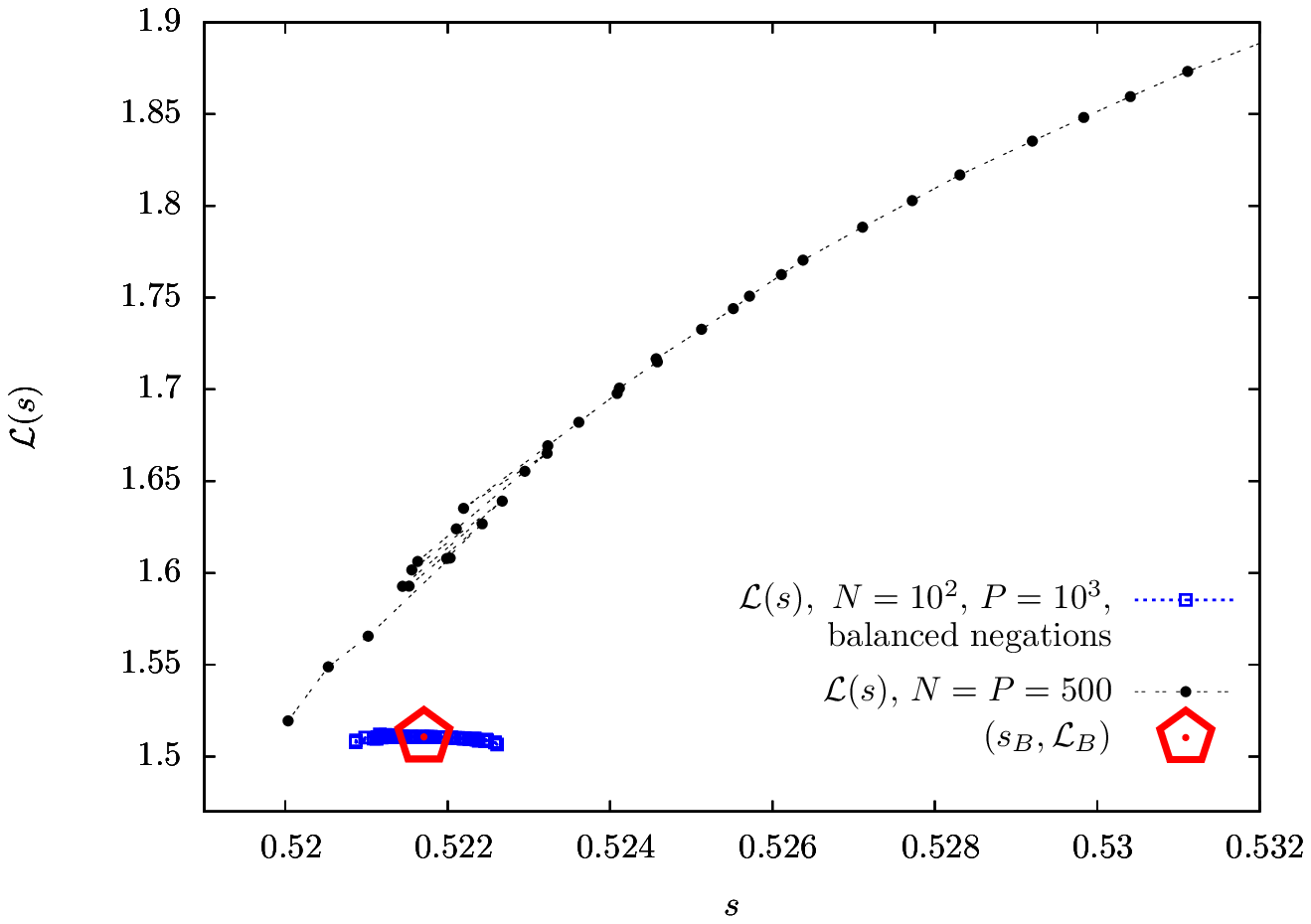}
\caption{
%{\bf Check the caption M OK} 
Left: The large deviation function $\mathcal{L}(s)$ vs. entropy $s$  computed by population dynamics on Poissonian graphs with random negation configurations, for $ K=3$, various values of the constraint density $\alpha$, and both positive and negative $x$. For $\alpha=1,2,3$ the part of the curve with $x<0$ has been computed with $N=P=300$. For $\alpha=1$ the part of the curve with $x>0$ reaches the right ending point, and has been computed with $N=P=500$. For $\alpha=2,3$ the part of the curve with positive $x$ has been computed with $N=P=700$, and it does not reach the right ending point, even larger $N$ and $P$ are needed to remove the noise from the data for very large positive $x$. For $x<0$, an unphysical branch (concave part of the curve) starts at $x_0 \approx -42,\, x_0 \approx -44,\, x_0 \approx -40$ for $\alpha=1,2,3$ respectively. Points indicate the values for balanced configurations of negations $(s_B,\mathcal{L}_B)$, and for configurations of negations where all the variables are non-negated $(s_R,\mathcal{L}_R)$.  \\
Right: Zoom at the large deviation function $\mathcal{L}(s)$ vs $s$ for $\alpha=1$ near to the low-entropy ending point for $N=P=500$. The data become noisy near to the low entropy ending point, larger graph and population sizes lead to an improvement. We plotted the data down to the lowest value of entropy $s$, hence the unphysical branch is not plotted.  The point indicates the value $(s_B,\mathcal{L}_B)$ for balanced configurations of negations. The blue data points is the large deviation function restricted to balanced configurations of negations for $N=10^2,\, P=10^3$. Notice the narrow range of entropies $s$ plotted and how little the lowest entropy we achieved differs from the balanced value. 
}
\label{fig3}
\end{figure}

In the most commonly considered ensemble of the random $K$-satisfiability problem the triple of variables appearing in each clause is chosen independently at random (avoiding repetitions). For large system sizes this procedure generates Poissonian degree distribution with mean $\alpha$. In this case every node has a different local neighborhood and hence the fixed point of Eqs.~(\ref{10}-\ref{11}) is not factorized, instead the distribution $S_{ia}(m_{ia},J_{ai})$ is different on every edge. We hence solve Eqs.~(\ref{10}-\ref{11}) by generating an instance of the problem (graph) of size $N$, associating one population of size $P$ with every directed edge and iterating following Eqs.~(\ref{10}-\ref{11}). This is more computationally involved and we are able to treat only modestly large $N$ and $P$, typically several hundreds. The resulting large deviation function $\mathcal{L}(s)$ is depicted in Fig.~\ref{fig3} for several values of constraint density $\alpha$.

%{\bf Check this paragraph. M OK} 
For low values of the constraint density, e.g. $\alpha=1$ in Fig.~\ref{fig3}, the location of the right (large entropy, least frustrated) ending point $(s_R,\mathcal{L}_R)$ corresponds, as in the case of random regular instances, to the value of Bethe entropy that is obtained if no negations are present in the instance ($J_{ia}=0$ for all $ia$), and $\mathcal{L}_{R}= (1-e^{-K\alpha})\log 2$ (corresponding to the number of negation-configurations where no variable is locally frustrated). For larger values of the  constraint density, e.g. $\alpha=2$ in Fig.~\ref{fig3}, the results from population sizes as large we were able to achieve are very noisy for large values of $x \approx 100$. We observed that the data are getting smoother as the population size is growing, however, not enough to be able to conclude from these data whether $(s_R,\mathcal{L}_R)$ is the right ending point. 

The part of the curve corresponding to a very large negative parameter $x$ does not converge to an ending point. Instead at some $x_0$ the large deviation function $\mathcal{L}(s)$ ceases to be concave, and an unphysical branch appears for $x<x_0$. This unphysical branch is not present on random regular instances with even degree, when the degree is odd the data for large negative $x$ are inconclusive in the sense that we might see a unphysical branch or only a numerical noise. We define the left ending point as the extreme of the physical branch $s_L=s(x_0)$, and $\mathcal{L}_L=\mathcal{L}(x_0)$. And we observe systematically that in the region of interest (say for $\alpha \ge 1$) the values $s_L$ and $\mathcal{L}_L$ are very close to the values corresponding to balanced instances $(s_B,\mathcal{L}_B)$. In balanced instances each variable is negated as many times as non-negated (for variables of odd degree the absolute value of the difference between the number of negations and non-negations is 1). In the thermodynamic limit the number of such balanced negation-configurations is 
\begin{equation}\label{23}
\mathcal{L}_{B}=\sum_{i=0}^ {\infty}\log\left(\begin{array}{c}
2i\\
i\end{array}\right)e^{-k\alpha}\frac{(k\alpha)^{2i}}{(2i)!}+\sum_{i=0}^ {\infty}\log\left[2\left(\begin{array}{c}
2i+1\\
i\end{array}\right)\right]e^{-k\alpha}\frac{(k\alpha)^{2i+1}}{(2i+1)!}.
\end{equation}

We made a number of attempts to obtain a value of entropy considerably smaller than the balanced entropy, $s<s_B$. First, we removed the leaves from the formula and balanced only the residual formula. This indeed leads to a lower value of entropy but for $\alpha>1$ the difference was less than $1\%$. We tested the population dynamics limited to the balanced negation configurations, i.e. we solved equations (\ref{10}-\ref{20}) where the sum over the negation-variables in Eqs. (\ref{11}) and (\ref{19}) was limited only to the balanced negation configurations.
The large deviation function $\mathcal{L}(s)$ obtained this way did not differ more than by $1\%$ from the value $(s_B,\mathcal{L}_{B})$, see Fig.~\ref{fig3} left. We also investigated the results of population dynamics over the SP equations and we were not able to find cases where the complexity would decrease by more than $1\%$ below the complexity value on the balanced instances. We also tried simulated annealing on the negation-variables using the Bethe entropy as the cost function, with the same result. All this makes us conclude that with at least $1\%$ of precision the satisfiability threshold for random adversarial SAT equals the satisfiability threshold of the balanced random ensemble.  

Let us hence summarize results about BP and SP for the random satisfiability problem with balanced configurations of negations. For $K=3$ the BP ceases to converge for $\alpha \ge 2.96$. SP starts to converge to a nontrivial fixed point for $\alpha>3.20$, and the complexity decreases to zero at 
\be 
            \alpha_B=3.399 \pm .001
\ee
this is hence the satisfiability threshold on the balanced random formulas. All our observations about the large deviation function suggest that the threshold for random adversarial satisfiability problem satisfies $\alpha_{a}>3.39$.

\section{Numerical results for AdSAT and large deviations}\label{sec_num}

In this final Section we compare theoretical predictions from the cavity method with numerical results.

\subsection{Numerical results for large deviations}\label{sub_dev}

First, we investigate numerically the number of  configurations of negations yielding a formula with a certain entropy of solutions. For one given random graph geometry of size $N$, we generate independently at random $I\gg 1$ different configurations of negations, and for each of them we count the number of solutions using a publicly available implementation of exact counting algorithm {\tt relsat} \cite{BayardoPehousek00}. We define probability $P_N(s)$ over the negation-configurations that the value of the entropy density was between $s$ and $s+\Delta s$, where $\Delta s$ is a binning interval that we will specify later. 

\begin{figure} 
\centering
\includegraphics[scale=0.7]{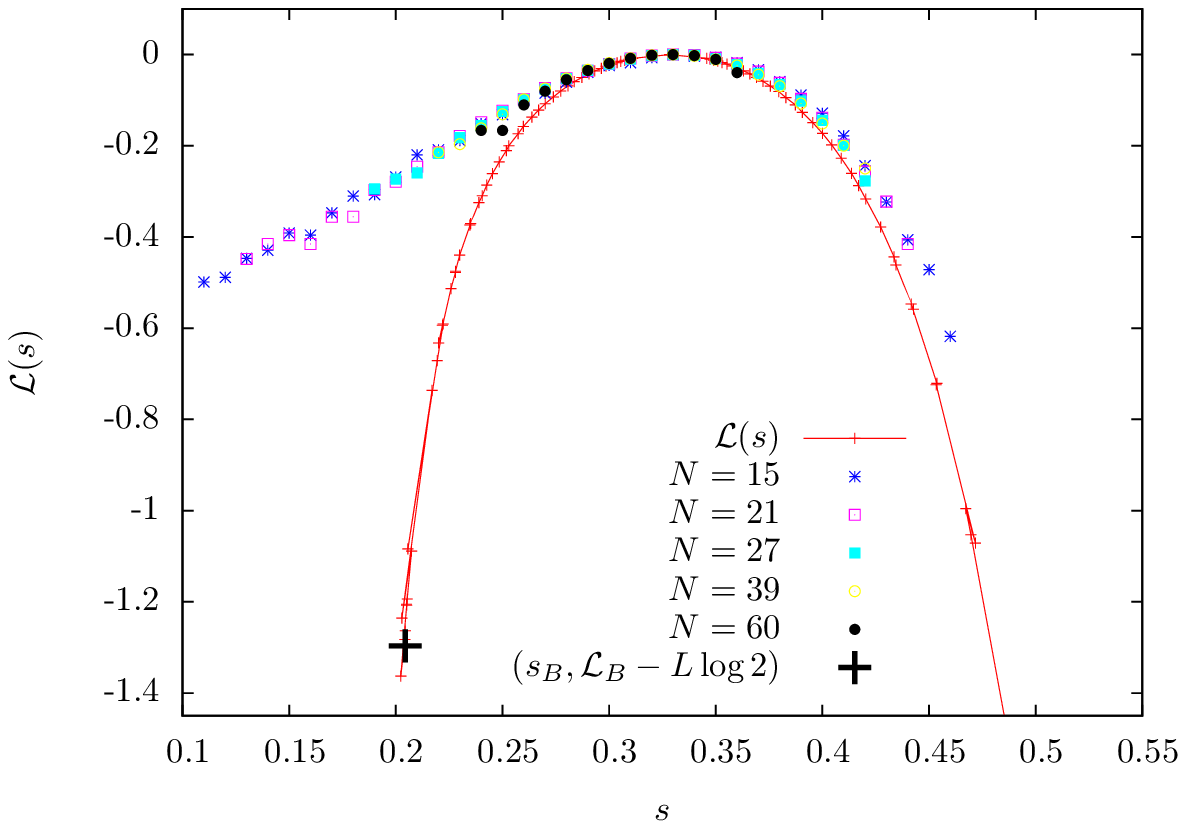}
\includegraphics[scale=0.7]{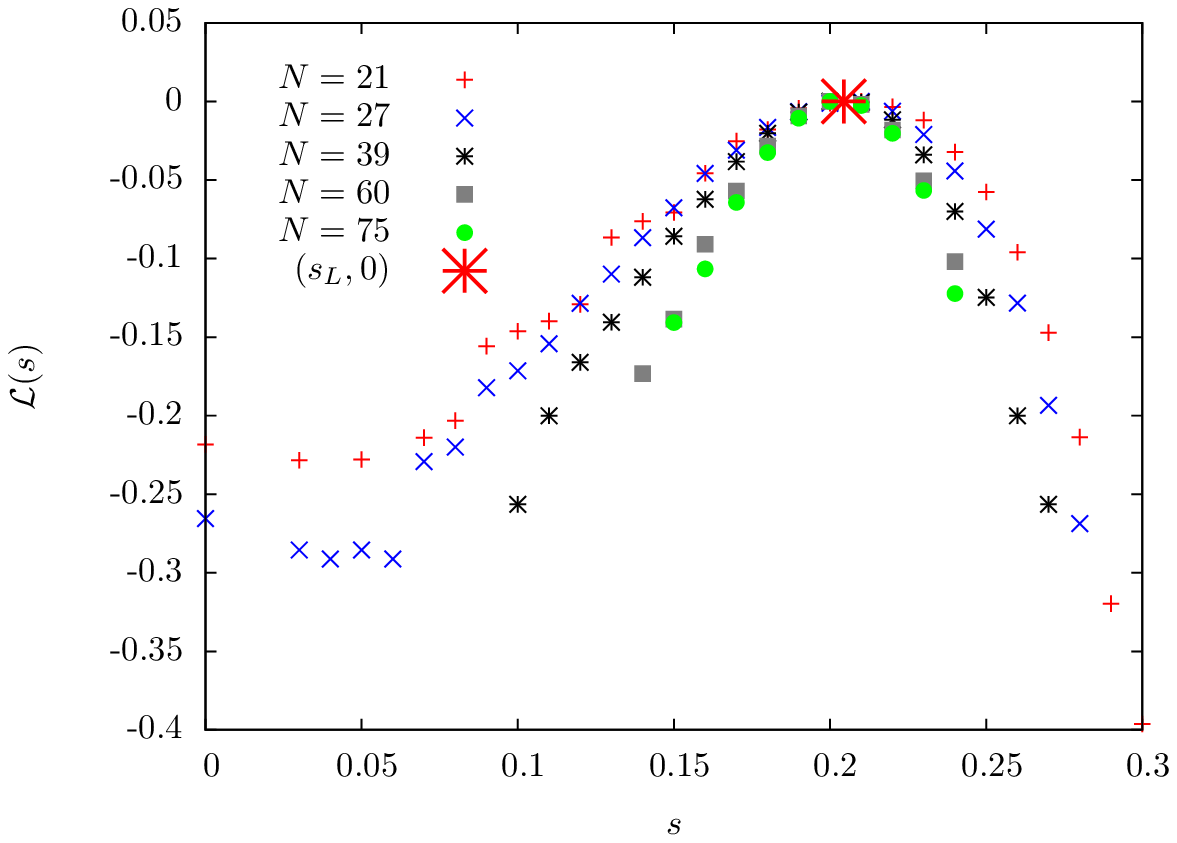}
\caption{Left: $\mathcal{L}_N(s)$ vs $s$  computed exactly for  $15 \leq N \leq 60$ with $N_s=10^ 5$ samples $\mathcal J$, and binning interval $\Delta s = 10^{-2}$,  and $\mathcal{L}(s)$ vs $s$  computed with the  population dynamics for both negative $x$ ($P=5 \cdot 10^ 4$) and positive $x$ ($P=2 \cdot 10^ 4$), with $K=3,\, L=8$. \\
Right: $\mathcal{L}_N(s) $ vs $s$ computed exactly for  $21 \leq N \leq 75,  N_s = 10^5$, binning interval $\Delta s=10^ {-2}$, and $K=3,\, L=8$. The curves don't superpose for $f<f_B$, so the left-large deviations are faster than exponential.}
\label{fig5}
\end{figure}

Following the assumption of exponentially large deviation function made in Eq.~(\ref{LL}), we define 
\be
  {\cal L}_{N}(s)   = \frac{1}{N} \left[ \log{P_N(s)} - \log{\max_s{P_N(s)}} \right].
\ee
The numerical result for $\mathcal{L}_N(s)$ is depicted in Fig.~\ref{fig5} left for $L=8$, and compared to the prediction of large deviation of the Bethe entropy from Section \ref{sec_res}. 
%{\bf  Check the rest of this paragraph M Made some infinitesimal changes} 
The agreement between the numerical data point and the theoretical prediction is not good in the low entropy region. One possibility is that this is due to pre-asymptotic effects, on the other hand this does not seem likely as the numerical curves seem to superpose nicely for different system sizes. Another possibility is that we neglected some replica symmetry breaking effects, note, however, that the large deviation calculation over survey propagation did not provide any non-trivial result. We hence leave this disagreement as an open problem.  

At this point we want to recall the result from BP we obtained on balanced regular instances with even degree (i.e. for instance $L=8$), in that case the BP fixed point was factorized and independent of the negation-configuration even for small graphs. Let us hence investigate the numerical results for entropy large deviations in this case. The data for ${\cal L}_{N}(s)$ are depicted in Fig.~\ref{fig5} right, recall from Table \ref{table_reg} that the maximum of the curve corresponds to  $s_B$ obtained with BP. The curves in Fig.~\ref{fig5} right clearly do not superpose for different system sizes. Instead the ${\cal L}_{N}(s)$ seems to be `closing'. From these data it is indeed plausible that in the  limit $N \rightarrow \infty$, ${\cal L}_{N}(s)$ converges to a delta function on the value of entropy $s=s_B$. 

Hence, data in Fig.~\ref{fig5} right  suggest that the probability that the entropy of a formula is different from the value predicted by BP is smaller than exponentially small. This makes us conclude that in a general case, the probability of having  an entropy outside of the interval $(s_L,s_R)$ is smaller than exponentially small (we remind that for the balanced negations and even degree $L$ regular graphs $s_L=s_R=s_B$). Hence in the thermodynamic limit there are almost surely no negation-configurations that would lead to a value of entropy outside the range $(s_L,s_R)$. 

Moreover, the large deviation function $\mathcal{L}_N(s)$ if asymptotically negative can be interpreted as a probability of generating a rare graph and configuration of negations having entropy $s$ \cite{Rivoire04}. Since there are of order $N^N$ regular graphs, and there is none or at least one graph with entropy $s\notin (s_L,s_R)$, we can have either $P_N(s)=0$ or 
\be \label{ineq}
  P_N(s) \ge e^{- c_1 N \log{N}}  
\ee
where $c_1$ is some positive constant. Consider now that there is $e^{N {\cal L}'}$ of configurations of negations (e.g. ${\cal L}'=K \alpha \log{2}$ if we consider all the negations-configurations, or  ${\cal L}'={\cal L}_B$ if we consider just the balanced negations-configurations). The fraction of graphs with configurations of negations leading to entropy $s\notin (s_L,s_R)$ has to be small only if
\be
        P_N(s) e^{N {\cal L}'} \ll 1. \label{bla}
\ee
If an equality holds in Eq. (\ref{ineq}) then Eq. (\ref{bla}) holds in the thermodynamic limit, $N\to \infty$. However, (\ref{bla}) does not have to hold for finite $N$ unless  
\be
        N \geq  N_c \equiv \exp \left( \frac{ {\cal L}' }{c_1} \right). \label{cross}
\ee 
Since $c_1$ can be considerably smaller than ${\cal L}'$, the crossover value of $N_c$ might be very large and out of reach for exact numerical methods. This justifies the presence of strong pre-asymptotic effects for system sizes treated in Fig.~\ref{fig5}.

\subsection{Strong finite size corrections for AdSAT threshold}\label{sub_sat}

We investigate numerically the AdSat threshold $\alpha_a$ by computing the probability (over random graph instances) $p_s$ that an adversary is not able to find a configuration of negations that makes the formula unsatisfiable. We do this in the regular instances because of reduced fluctuations that arise because of the randomness of the graph. 

\begin{figure} 
\centering
\includegraphics[scale=0.7]{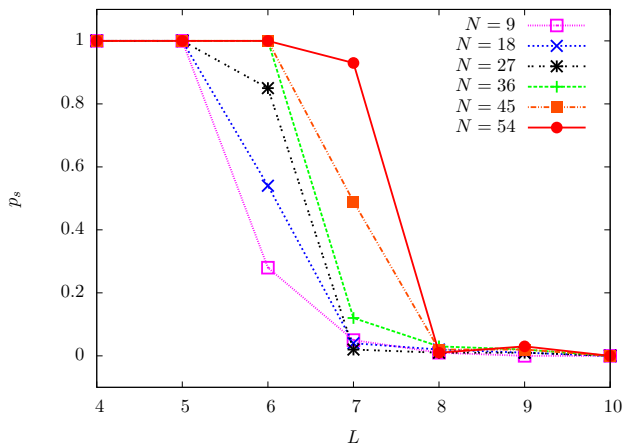}
\caption{Probability  $p_s$ that using the simulated annealing algorithm described in the text we did not find any unsatisfiable configuration of negations for $L$-regular instances, as a function of $L$, for different systems sizes $9\leq N \leq 54$. We used the  annealing rate $r=1.1$, and number of instances $I=100$.}
\label{fig11}
\end{figure}

We generate $I\gg 1$ regular instances for each value of the degree $L$ and for each size $N$. Then for each instance we use simulated annealing on the negation-variables in order to minimize the number of solutions, we monitor if an unsatisfiable formula is generated or not. This general strategy for AdSAT was suggested by \cite{Scardicchio_com}. In particular, we introduce an inverse temperature $\beta$. Initially we set $\beta=1$. We choose randomly one of the negation-variables, $J_{ia}$, and attempt to flip it, i.e. to set $J_{ia} \rightarrow 1- J_{ia}$. Denoting by $\mathcal J'$ the configuration  of negations after this flip, we accept the  flip with probability $\max \{ 1, e^{ \beta \left(  s_{\mathcal J'} - s_{\mathcal J}  \right) } \}$. The entropy $s_{\mathcal J}$ is  computed  exactly with a publicly available implementation of exact exhaustive search algorithm {\tt relsat} \cite{BayardoPehousek00}. This algorithm has an exponential running time in the size of the system, limiting us to very small system sizes. Attempting for $N$ negation flips is one Monte-Carlo (MC) step. Every 10 MC steps we multiply the inverse temperature by a rate factor $r>1$. We keep track of the so far minimal value of entropy $s_{\rm min}$ and the index $n_0$ of the MC step in which it was first found. The algorithm stops if either an unsatisfiable instance was encountered or no further decrease in the value of entropy $s_{\rm min}$ has occurred in the last $9 \cdot n_0+50$ MC steps. The probability $p_s$ plotted in Fig.~\ref{fig11} is then given by the fraction of cases in which an unsatisfiable instance was not found. 

There is of course no guarantee that our algorithm found the actual minimal possible entropy. So, strictly speaking, any result for the  satisfiability threshold derived from the data for $p_{s}$ is only an upper bound to the true threshold. However, given the strictness of our stopping condition we have a reasonable confidence that our results are very close to the exact result. Fig.~\ref{fig11} depicts the fraction of regular instances of size $N$ where we were unable to find a configuration of negations that would make the formula unsatisfiable. 

On a first sight numerical data in Fig.~\ref{fig11} do not agree with our theoretical predictions. Indeed, we predicted that unsatisfiable configurations of negations exist only for $L\ge 11$, whereas, on the system sizes our simulated annealing algorithm was able to treat, we find unsatisfiable negation-configurations for a large fraction of graphs with $L\ge 8$. 

On a second sight, however, we see in Fig.~\ref{fig11} that for $L=6$ and $L=7$  there are very strong finite size corrections to $p_{s}$. Indeed, for $L=6$ and size $N=9$ we find that roughly $3/4$ of the instances can be made unsatisfiable, whereas for $N=36$ none of the $I=100$ instances that we tried can be made  unsatisfiable. Similarly for $L=7$ and size $N=36$ we find that most of the $I=100$ instances can be made  unsatisfiable, whereas for $N=54$ almost none of them. If this trend continues it is perfectly plausible that in the $N\to \infty$ limit even for $L=10$ the adversary is never successful. These results, with agreement with conclusions of the previous section, suggest very strong pre-asymptotic effects in the AdSAT problem. The strength of the finite size corrections hence poses a challenge to numerical verifications of our cavity method asymptotic predictions.  

On the other hand, the scaling argument presented in Eq. (\ref{cross}) suggests that the system sizes at which the asymptotic behavior starts to be dominant might be quite large (perhaps thousands or more), this is in particular true in the vicinity of the satisfiability threshold. Hence in the AdSAT problem, and likely also in other adversarial optimization problems, it is particularly important to develop techniques that predict the pre-asymptotic behavior and the finite size corrections. We saw from the results on random regular instances with even degree that BP predicts the same Bethe entropy for all balanced negation-configurations independently of the system size, hence the methods for analysis of finite size corrections and pre-asymptotic effects will have to go beyond the assumptions of the cavity method. On the other hand, analysis of the cases where the BP fixed point is factorized might be a good playground for development of such techniques. 

\section{Discussion and conclusions}\label{sec_concl}

In this paper we studied the adversarial optimization problem and concluded that the most frustrated instances of random $K$-SAT are very close to the ones with balanced configurations of negations. For random regular 3-SAT instances this leads to a threshold $L=11$, starting from which the adversary is able to find unsatisfiable configurations of negations (compare to $L=14$ for the ordinary random regular 3-SAT). For the canonical (Poissonian) adversarial 3-SAT this leads to $\alpha_{a}=3.39(1)$ (compare to $\alpha_c=4.2667$ for the ordinary random 3-SAT).  

This result is rather uninteresting from the algorithmic point of view, as balancing negations is an easy problem. However, the same method we used here can be applied to more interesting situations, for instance the quantified SAT problem. Recall also that the adversarial satisfiability problem was suggested as a problem interpolation between random SAT and random quantum SAT. Note, however, that our study leads to the conclusion that the adversary SAT is much closer to the classical random SAT than to the quantum SAT. Note that in the large $K$ limit the random satisfiability threshold scales as $\alpha \approx 2^K\log{2}$, the same scaling holds for the threshold in the random $K$-SAT with balanced negations, since at large $K$ at this value of $\alpha$ the degree of variables is so large that the difference between the Poissonian distribution of the number of non-negated variables and balanced negation-configurations does not play any role in the leading order in $K$. On the other hand, the satisfiability threshold of the quantum SAT was upper-bounded by $2^K\log{2}/2$ \cite{BravyiMoore09,LaumannLauchli10}, hence quantum effect must be responsible for this drastic decrease of the threshold value.

We obtained our results by studying the large deviations of the entropy in the ordinary random $K$-SAT. In particular, an approach leading to equations very similar to the $1$RSB equations leads to the calculation of the large deviations  in the case where rare instances are exponentially rare. Exponentially large deviations are common in statistical physics. In some cases, see e.g. \cite{MonthusGarel10,ParisiRizzo08,ParisiRizzo09,ParisiRizzo10}, the large deviations are rarer than exponentially rare. In our study this arises for instance for the regular random $K$-SAT with balanced negations and even degree. In cases the large deviation function decays faster than exponentially with the system size extremely strong finite size corrections and pre-asymptotic effects can be induced as we argued in Section \ref{sec_num}, where we presented numerical studies of the large deviations and of the satisfiability threshold. Interestingly, methods based on the standard cavity method are not  straightforwardly applicable to study the related finite size correction and pre-asymptotic behavior. It stays a theoretical challenge to find out how to describe analytically and algorithmically these pre-asymptotic effects that might be crucial for solving some industrial instances of adversarial optimization problems. 

\section{Acknowledgment}

We thank Cris Moore for introducing us to the adversarial SAT problem, Antonello Scardicchio for very helpful discussions, and Guilhem Semerjian for very helpful discussions and very useful comments about a preliminary version of the manuscript. We also acknowledge support from the D. I. computational center of University \textit{Paris Sud}. 

\section*{References}

\bibliographystyle{unsrt}
\bibliography{myentries}

\end{document}